\documentclass[a4paper,fleqn,usenatbib]{mnras}

\usepackage{graphicx}
\usepackage{amsmath}
\usepackage{amssymb}
\usepackage{newtxtext,newtxmath}
\usepackage[T1]{fontenc}
\usepackage{ae,aecompl}

\title[X-ray and radio studies of SNR CTB 37B]{X-ray and radio studies of SNR~CTB~37B hosting the magnetar CXOU~J171405.7$-$381031}
\author[Blumer et al.]{Harsha Blumer$^{1,2}$, Samar Safi-Harb$^{3}$, Roland Kothes$^4$, Adam Rogers$^5$, Eric V. Gotthelf$^6$\\
$^{1}$ Department of Physics and Astronomy, West Virginia University, Morgantown, WV 26506, USA; harsha.blumer@mail.wvu.edu\\
$^{2}$ Center for Gravitational Waves and Cosmology, West Virginia University, Chestnut Ridge Research Building, Morgantown, WV 26505, USA \\
$^{3}$Department of Physics \& Astronomy, University of Manitoba, Winnipeg, MB R3T 2N2, Canada; Samar.Safi-Harb@umanitoba.ca\\
$^{4}$Dominion Radio Astrophysical Observatory, National Research Council Herzberg, P. O. Box 248, Penticton, BC V2A 6J9, Canada; roland.kothes@nrc-cnrc.gc.ca \\
$^5$ Department of Physics \& Astronomy, Brandon University, Brandon, Manitoba, R7A 6A9, Canada;  rogersa@brandonu.ca\\
$^6$Columbia Astrophysics Laboratory, Columbia University, 550 West 120th Street, New York, NY 10027-6601, USA; eric@astro.columbia.edu}

\begin{document}
\label{firstpage}
\pagerange{\pageref{firstpage}--\pageref{lastpage}}

\maketitle

\begin{abstract} 
We present a \textit{Chandra} and \textit{XMM-Newton} study of the supernova remnant (SNR) CTB 37B, along with archival radio observations. In radio wavelengths, the SNR CTB 37B is an incomplete shell showing bright emission from the eastern side while the X-ray morphology shows diffuse emission from regions surrounding the magnetar CXOU J171405.7--381031. We used archival HI absorption measurements to constrain the distance to the remnant and obtain $D$~=~9.8$\pm$1.5~kpc. The X-ray spectrum of the remnant is described by a thermal model in the 1--5 keV energy range, with a temperature of $kT$~=~1.3$\pm$0.1~keV. The abundances from the spectral fits are consistent with being solar or sub-solar.  A small region of diffuse emission is seen to the southern side of the remnant, best fitted by a nonthermal spectrum with an unusually hard photon index of $\Gamma$~=~1.3$\pm$0.3. Assuming a distance of 9.8~kpc to the SNR, we infer a shock velocity of $V_s$~=~915$\pm$70~km~s$^{-1}$ and an explosion energy of $E$~=~(1.8$\pm$0.6)$\times$10$^{50}$~ergs. The overall imaging and spectral properties of CTB~37B favor the interpretation of a young SNR ($\lesssim$6200 yr old) propagating in a low-density medium, under the assumption of a Sedov evolutionary phase. 
\end{abstract}

\begin{keywords}
ISM: individual (CTB 37B) -- supernova remnants -- X-rays: ISM -- pulsars: individual (CXOU J171405.7--381031)
\end{keywords}

\section{Introduction}

The CTB 37 complex is one of the more active regions in our Galaxy with the presence of numerous supernova remnants (SNRs), molecular clouds, starburst activities, and hosts several sources of very high-energy (VHE) gamma-rays. The complex includes the three radio supernova remnants (SNRs) CTB 37A,  CTB 37B, and G348.5--0.0 (Clark et al. 1975; Milne et al. 1979; Kassim et al. 1991). The discovery of TeV gamma-ray source J1713--381 (Aharonian et al. 2006) using the High Energy Stereoscopic System (H.E.S.S.), coincident with SNR CTB 37B, initiated X-ray observations to search for its X-ray counterpart. A galactic plane survey with Advanced Satellite for Cosmology and Astrophysics (ASCA) detected a part of CTB~37B, where Yamauchi et al. (2008) showed that the SNR spectrum was a mixture of a nonthermal powerlaw (photon index $\Gamma$~$\sim$~4.1) and an optically thin thermal plasma (temperature $kT$~$\sim$~1.6~keV) emission.  Using Chandra data, Aharonian et al. (2008) identified the point source CXOU J171405.7--38103 coincident with the radio shell of CTB~37B. This potential counterpart to the SNR displays a soft nonthermal ($\Gamma$~$\approx$~3.3) spectrum and was subsequently identified as a 3.82~s pulsar, a likely magnetar (Halpern \& Gotthelf 2010a). Additional {\it XMM-Newton} observations confirmed the magnetar nature of the source, whose rapid spin-down rate implies a magnetic field strength of $B$~=~4.8$\times$10$^{14}$~G and spin-down power of $\dot{E}$~=~4.2$\times$10$^{34}$~ergs~s$^{-1}$ (Sato et al. 2010; Halpern \& Gotthelf 2010b). The \textit{Chandra} data also revealed thermal emission from $\sim$4$\arcmin$ region close to the radio shell of CTB 37B and implied an SNR age of $\sim$5000 years and an ambient gas density of $\sim$0.5~cm$^{-3}$ (Aharonian et al. 2008).  Subsequently, a Suzaku study of the SNR CTB 37B was performed and the spectra revealed a thermal component of $kT$ = 0.9$\pm$0.2~keV and a nonthermal emission of $\Gamma$~$\sim$~1.5 from the southern region (Nakamura et al. 2009). These authors also suggested an SNR age of 650$^{+250}_{-300}$ yr and a pre-shock electron density of 0.4$\pm$0.1~cm$^{-3}$.  

Owing to the discrepancy in the SNR age estimates (ranging from 650 yr to 5000 yr) and the lack of well constrained spectral parameters in the above mentioned studies, we perform a detailed imaging and spectroscopic analysis of the remnant using all available archival \textit{Chandra} and \textit{XMM-Newton} data to investigate the remnant's morphology, spectral parameters, and supernova explosion properties. A dedicated study of the magnetar will be published elsewhere (Gotthelf et al., in preparation). Archival radio data using the Molonglo Observatory Synthesis Telescope (MOST) are used to compare the radio and X-ray morphologies of the remnant, and to determine the distance to the SNR. The structure of the paper is as follows: Section 2 describes the \textit{XMM-Newton} and \textit{Chandra} observations and data reduction. The X-ray imaging and spectral analyses are presented in Sections 3 and 4, respectively. In Section 5, we discuss the derived properties and in Section 6, we present our conclusions.

\section{Observations \& Data reduction}
\label{2}

\subsection{XMM-Newton}
\label{2.1}

We use archival \textit{XMM-Newton} observations of the SNR CTB 37B, obtained with the European Photon Imaging Camera (EPIC) pn (Struder et al. 2001) and MOS (Turner et al. 2001) cameras covering the 0.2--12 keV energy range.  Here, we use only the data collected with the EPIC-pn and MOS cameras when the observations were carried out in Full Frame mode with a time resolution of 73.4 ms and 2.6~s, respectively. The data were analyzed with the \textit{XMM-Newton} Science Analysis System (SAS) software version 17.0.0\footnote{See http://xmm.esac.esa.int/sas/} and the most recent calibration files. Data from the contaminating background flares were excluded from our analysis. The event files were created from observational data files (ODFs) using the SAS tasks \textit{epchain} and \textit{emchain}. The events were then filtered to retain only patterns 0 to 4 for the pn data (0.2--15.0 keV) and patterns 0 to 12 for the MOS data (0.2--12.0 keV). The data were screened to remove spurious events and time intervals with heavy proton flaring by inspecting the light curves for each instrument separately at energies above 10 keV. The resulting total effective exposure times for the MOS1/2 and pn cameras are shown in Table~1.

\subsection{\textit{Chandra}}
\label{2.2}

The region around CTB 37B was also observed with the Advanced CCD Imaging Spectrometer (ACIS-I) onboard the \textit{Chandra} X-ray observatory on 2007 February 2 (ObsID: 6692) for a total exposure time of 25 ks. The data are configured in the VFAINT timed exposure mode with a CCD frame readout time of 3.2~s. We reduced the data using the standard \textit{Chandra} Interactive Analysis of Observations (CIAO) version 4.10\footnote{http://cxc.harvard.edu/ciao.} routines. The CIAO tool \textit{chandra\_repro} was applied to perform initial processing and obtain the event 2 file. The bad grades were filtered out and good time intervals were reserved. The resulting effective exposure time after data processing is summarized in Table 1, which was used to perform an imaging and spectral analysis. 

\begin{table*}
\caption{Observation log of SNR CTB 37B}
\begin{tabular}{l l l l l l}
\hline\hline Satellite & ObsID & Date of  &  Detector & Total & Effective  \\
& & observation & & Exposure (ks) & Exposure (ks)\\
\hline\hline
\textit{Chandra} & 6692 & 2 February 2007 & ACIS-I &  26 & 24.8 \\
\textit{XMM-Newton} &  0606020101 & 17 March 2010 & MOS1 & 120 & 93.2\\
& & & MOS2 & 120 & 94.2 \\
&  & & pn & 120 & 78.9 \\
& 0670330101 & 13 March 2012 & pn & 17 & 8.0 \\
& 0790870201 & 23 September 2016 & pn & 30 & 23.2 \\
& 0790870301 & 22 February 2017 & pn & 23 & 16.2 \\
\hline
\end{tabular}
\end{table*}

\section{Imaging analysis}
\label{3}

We performed an imaging analysis to investigate the SNR morphology in X-rays and radio. Figure 1 shows a truecolor image of CTB~37B (not corrected for the background or vignetting) with the radio emission in red, soft (0.5--2.0 keV) X-ray emission in green, and hard (2.0--8.0 keV) X-ray emission in blue. The {\it XMM-Newton} MOS 1 and 2 data were first divided into individual images in the soft (0.5--2.0 keV) and hard (2.0--8.0 keV) energy bands, and smoothed using a Gaussian function with $\sigma$~=~10$\arcsec$.  The archival radio image of the remnant, with a Gaussian smoothing of $\sigma$ = 10$\arcsec$, was obtained with MOST as part of the Sydney University Molonglo Sky Survey (SUMSS; Mauch et al. 2003). The radio coordinates of the remnant is given by $\alpha_{J2000}$~=~17$^{h}$13$^{m}$55$^{s}$ and $\delta_{J2000}$~=~$-$38$^{\circ}$11$\arcmin$00$\arcsec$ (Green 2009). 

\begin{figure}
\includegraphics[width=0.5\textwidth]{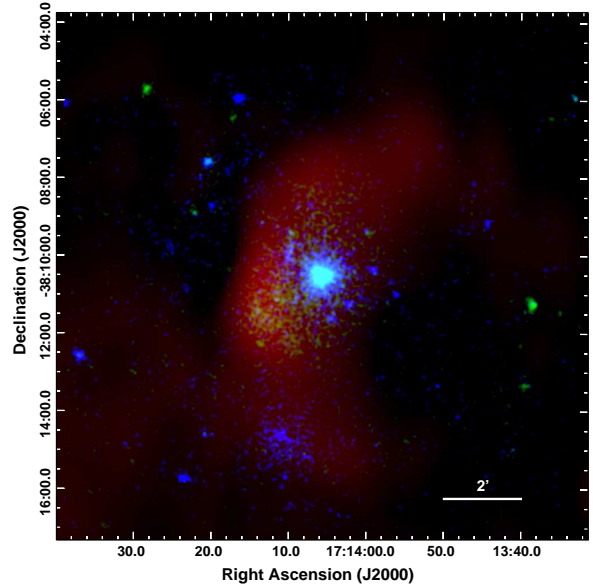}
 \caption{Truecolor image of SNR CTB~37B showing the radio and X-ray emission. The radio emission, shown in red, is obtained with MOST as part of the Sydney University Molonglo Sky Survey. The soft (0.5--2.0 keV) and hard (2.0--8.0 keV) X-ray emission with \textit{XMM-Newton} are shown by the colors green and blue, respectively. The images are shown on a logarithmic scale with a Gaussian smoothing of  $\sigma$ = 10$\arcsec$. Background subtraction and vignetting correction were not applied. See Section 3 for details.} 
 \end{figure}

In the radio band, CTB 37B has been classified as a partially limb-brightened shell-type SNR (Kassim et al. 1991). The radio image shows bright emission from the eastern side of the remnant while the western side appears nearly featureless. The \textit{XMM-Newton} image, however, shows that the emission is confined to an elliptical region surrounding the magnetar and elongated towards the southern side.  The X-ray image also reveals soft emission from the center and hard emission from the southern region (see Table~3; regions~1 and 2). Although a clearly defined shell is not evident in the X-ray image, the radio shell appears to overlap with the bright X-ray emission on the eastern side. Green (2009) had catalogued the radio emission to be 17$\arcmin$ in size, the average of 16$\arcmin$ $\times$ 18$\arcmin$ shown by the MOST image in Kassim et al. (1991), which included a faint ``bridge'' emission extending to the south-east.  Kassim et al. (1991) discussed this bridge emission with VLA data and argued that it is a blowout emission associated with CTB 37B.  However, this faint bridge emission is not seen in the X-ray images. Aharonian et al. (2008) estimate the extent of the SNR shell as $\sim$5$\arcmin$.1 based on the partial radio shell. Since the size and shape of the remnant is unknown, we here assume a radius of 5$\arcmin$.1 for the derivation of SNR properties.   

\section{Spectral analysis}
\label{4}

The spectral analysis was performed using the X-ray spectral fitting package, {\tt XSPEC} version 12.8.2\footnote{http://xspec.gsfc.nasa.gov.}, combining the \textit{Chandra} and \textit{XMM-Newton} data. The point 
sources in the extraction regions were identified and excluded from the source spectra using the standard methods documented 
for each mission.

Based on the results from our imaging analysis, we defined three main regions from both \textit{Chandra} and \textit{XMM-Newton} observations for spectroscopy -- (a) Region 1: diffuse emission seen towards the southern side of the remnant, which also overlaps with the radio shell on the eastern side, (b) Region 2: central diffuse emission around the magnetar, and (c) Region 3: global SNR of $\sim$5$\arcmin$.1 radius as estimated by Aharonian et al. (2008) to investigate the properties of the SNR shell. The regions are shown in Figure 2 (left) and their properties are summarized in Table 2. We excluded a region of $\sim$10$\arcsec$ and $\sim$35$\arcsec$ (accounting to $>$90\% encircled energy) around the magnetar from the \textit{Chandra} and \textit{XMM-Newton} data, respectively, to avoid any possible dust scattering or contamination from the magnetar. The background regions were selected from nearby source-free regions from the same CCD chip as the source and for the global SNR, we extracted an annular background region extending from 5$\arcmin$.2--7$\arcmin$.2 (as shown in Figure~2 (left)). For regions 1, 2, and 3, the \textit{XMM-Newton} spectra were grouped by a minimum of 25, 25, and 50 counts per bin and the \textit{Chandra} spectra were grouped by a minimum of 20, 25, and 25 counts per bin, respectively. The errors quoted are at the 90$\%$ confidence level.  Unless otherwise noted, spectra were fitted in the 1--5 keV range due to the high column
density and the Galactic background, at the lower and higher range, respectively.

\begin{figure*}
\includegraphics[width=\textwidth]{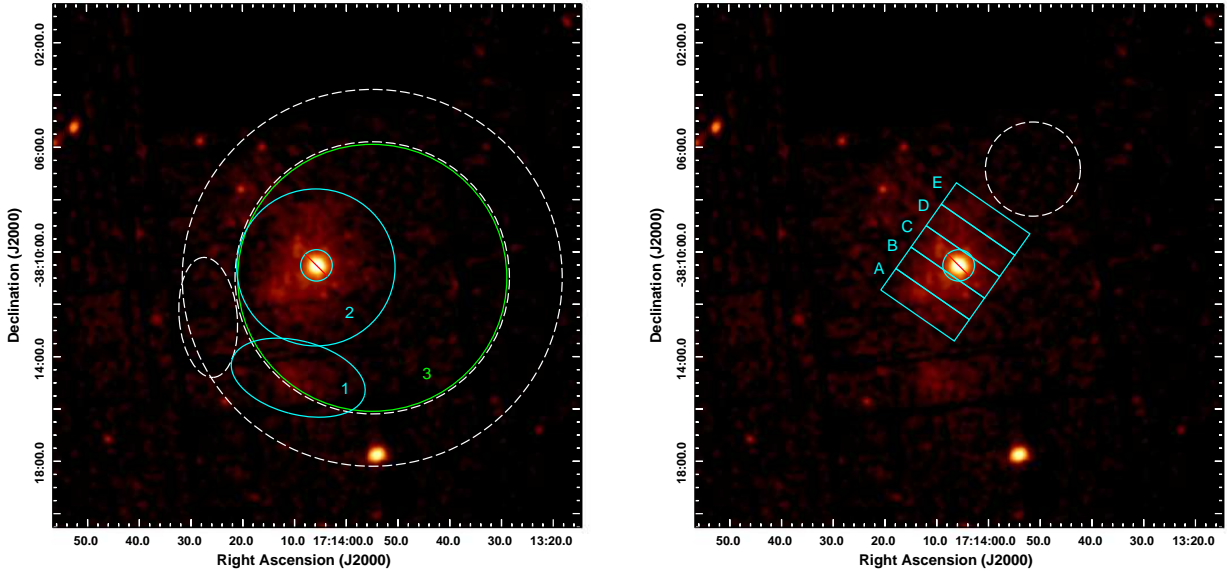}
\caption{Left: \textit{XMM-Newton}-MOS intensity image of CTB 37B (in logarithmic scale) smoothed using a Gaussian scale of $\sigma$ = 10$\arcsec$ overlaid with the regions selected for spectral extraction. The small scale regions 1 and 2 are shown in cyan whereas the global SNR (5$\arcmin$.1) is shown in green. Background extraction regions in white are denoted by a dotted ellipse, annulus, and circle. Right: Same as left, with raster scan boxes (in cyan) over the central soft diffuse emission (Region 2), enumerated as A to E in Table~4, from south to north. The image contrast has been reduced to display the regions and numbers.  All the point sources and central source circle were excluded in these cases. North is up and east is to the left.}
\end{figure*}

\begin{table}
\caption{Regions of diffuse emission extracted from CTB 37B using \textit{Chandra} and \textit{XMM-Newton}}
\begin{tabular}{l l l l l}
\hline\hline
Region (Number) & Right Ascension & Declination & Size \\
& h~m~s (J2000) & d m s (J2000) & \\
\hline
Region 1 & 17:14:09.282 & $-$38:14:48.92 & 2$\arcmin$.6$\times$1$\arcmin$.4  \\
Region 2 & 17:14:05.869 & $-$38:10:36.18 & 3$\arcmin$.0 \\
Region 3 & 17:13:55 & $-$38:11:00  & 5$\arcmin$.1  \\
 \hline
\end{tabular}

\end{table}

The diffuse emission within the remnant was investigated using different thermal and nonthermal models. For the thermal emission, we used both the collisional ionization equilibrium (CIE) models such as VMEKAL (Mewe et al. 1985; Liedahl et al. 1990) and VRAYMOND (Raymond \& Smith 1977), for describing plasma that has reached ionization equilibrium, and non-equilibrium ionization models (NEI) such as PSHOCK, NEI, VPSHOCK, and VSEDOV (Borkowski et al. 2001) for describing the plasma in young SNRs. All the models included the Tuebingen-Boulder ISM absorption model ($tbabs$ in XSPEC), with abundances set to those from Wilm et al. (2000). The spectra were first fit by treating the column density, temperature, and normalization as free parameters. The abundances of all elements were initially fixed at their solar values given by Anders \& Grevesse (1989) and varied as necessary to constrain the individual abundances. However, the abundances of H, He, C, and N were kept frozen to their solar values and Ni was tied to Fe throughout the spectral fitting.  In order to explore the high-energy continuum arising from any nonthermal emission, we used a powerlaw (PL) model. The CIE models resulted in a large reduced chi-squared $\chi^2_{\nu}$ $>$ 2 ($\nu$ being the number of degrees of freedom), which led us to rule out the CIE models. Subsequently, we fitted the spectra using NEI models which yielded better results. In Table~3, we list the best fit spectral parameters for all regions.

Region~1 (south diffuse emission): We considered both thermal and nonthermal one-component models to fit region 1 spectrum in the 1--8 keV energy range. The thermal models gave an unrealistically high temperature ($kT$ $\sim$ 30~keV with bremsstrahlung model) and posed difficulty in constraining the errors while a PL model gave a good fit for the southern diffuse emission. Figure~3 shows the best fit PL spectrum for region 1.

Region~2 (central diffuse emission): This region clearly revealed the presence of emission lines from Mg, Si, and S and hence, we examined the spectra using thermal models. We obtained nearly solar values for all the elements and the fit did not improve significantly by letting Mg, Si, and S abundances to vary. A VPSHOCK model best described the emission from the central diffuse emission region (Figure~4). Adding a second thermal or nonthermal component to the spectral model was not justified statistically in the fits.

We also checked for any spatial variation associated with the soft thermal emission (region 2) using the 2010 deep imaging \textit{XMM-Newton} observation. We partitioned the diffuse emission into a series of 5 adjacent $1\farcm0 \times 3\farcm4$ strips as denoted in Figure~2 (right) and fitted with an absorbed NEI model.  The background was selected from a nearby circular region ($r$ $=$ $2^{\prime}$). Invoking a variable abundance NEI model does not result in an improved fit and hence, the abundances were fixed to their solar values for the spatial analysis. Table~4 summarizes the result of spectral fits to these strips in which we consider the column density either independent or linked.  We find marginal change in the column density across this region suggesting that the absorption is mostly interstellar. The temperature ($kT$) appears to cool away from the magnetar, a possibly interesting result, suggesting evidence that they are related.

Region~3 (global SNR): We performed the spectral fitting for global SNR spectra using VPSHOCK and VSEDOV models, letting Mg, Si, and S to vary.  The abundances were consistent with being solar or sub-solar. Addition of a second thermal or nonthermal component did not improve the fits. Figure~5 shows the spectrum extracted from the whole SNR fitted with a VSEDOV model.

\begin{table*}
\center
\caption{Best fit spectral parameters of CTB 37B using \textit{Chandra} and \textit{XMM-Newton}}
\begin{tabular}{l l l l l}
\hline\hline 
Parameter & Region 1 & Region 2 & Region 3  \\
\cline{4-5}
& PL & VPSHOCK & VPSHOCK & VSEDOV\\
\hline
 
 $ N_{H}$ ($10^{22}$ cm$^{-2}$) & 3.1$_{-0.8}^{+0.9}$ & 4.3$_{-0.1}^{+0.2}$ & 4.3$\pm$0.3 & 4.2$\pm$0.2\\
  
  $\Gamma$ & 1.3$\pm$0.3 & - & - & -  \\

  kT (keV) & - & 1.4$\pm$0.1 & 1.8$_{-0.4}^{+0.2}$ & 1.3$\pm$0.1 \\

 Mg  & -  & 0.9$\pm$0.2 & 0.9$\pm$0.2 & 1.2$\pm$0.5 \\

 Si  & - & 1.2$\pm$0.1 & 1.1$\pm$0.1 & 1.0$\pm$0.5 \\

 S  & - & 0.8$_{-0.1}^{+0.2}$ & 0.8$\pm$0.2 & 0.9$_{-0.2}^{+0.1}$  \\

 $n_et$ (cm$^{-3}$s) & - & 3.8$_{-0.8}^{+1.1}\times10^{10}$ & 3.0$_{-0.6}^{+0.8}\times10^{10}$ & 5.8$_{-0.8}^{+1.0}\times10^{10}$ \\

$f_{unabs}$ (VP) & - & 5.7$_{-0.5}^{+0.7}\times10^{-11}$ & 7.2$_{-1.0}^{+1.1}\times10^{-11}$ & 9.8$_{-1.1}^{+0.9}\times10^{-11}$ \\

$f_{unabs}$ (PL) & 5.9$_{-1.5}^{+2.1}\times10^{-13}$ & - & -  & - \\

 $\chi_{\nu}^2 (dof)$ & 1.09 (863) &  1.06 (1499) & 1.06 (1905) & 1.07 (1904) \\
\hline
\end{tabular}

Errors are 2 $\sigma$ uncertainties.  The unabsorbed fluxes ($f_{unabs}$) quoted are in units of ergs cm$^{-2}$ s$^{-1}$ for the energy range 1--8 keV.
\end{table*}

\begin{table}
\caption{Spatially resolved NEI fits to Region 2}
\begin{tabular}{l l l l}
\hline\hline 
Slice & $ N_{H}$ & kT  & $n_et$  \\
& ($\times$10$^{22}$~cm$^{-2}$) & (keV) & ($\times$10$^{10}$~s~cm$^{-3}$)  \\
\hline
A & $4.9^{+0.5}_{-0.4}$ & $1.0^{+0.2}_{-0.2}$ & $5.5^{+7.1}_{-2.3}$ \\
 B & $4.7^{+0.4}_{-0.3}$ & $1.3^{+0.2}_{-0.2}$ & $2.2^{+0.8}_{-0.5}$ \\
 C & $4.4^{+0.4}_{-0.4}$ & $2.0^{+0.8}_{-0.4}$ & $0.7^{+0.4}_{-0.2}$ \\
 D & $4.3^{+0.4}_{-0.4}$ & $1.5^{+0.4}_{-0.3}$ & $0.7^{+0.3}_{-0.2}$ \\
 E & $4.2^{+0.6}_{-0.6}$ & $1.3^{+0.6}_{-0.3}$ & $2.0^{+1.9}_{-0.8}$ \\
\hline
A & $4.3$ (fixed)  & $1.1^{+0.1}_{-0.1}$ & $5.0^{+2.6}_{-1.8}$ \\
 B & $4.3$ (fixed)  & $1.4^{+0.1}_{-0.1}$ & $2.2^{+0.7}_{-0.4}$ \\
 C & $4.3$ (fixed)  & $1.9^{+0.4}_{-0.3}$ & $0.7^{+0.3}_{-0.2}$ \\
 D & $4.3$ (fixed)  & $1.4^{+0.2}_{-0.2}$ & $0.8^{+0.2}_{-0.2}$ \\
 E & $4.3$ (fixed)   & $1.1^{+0.2}_{-0.2}$ & $2.0^{+1.5}_{-0.7}$ \\
\hline
\end{tabular}   
                                                 
Spectral fits to the 2010 {\it XMM-Newton} MOS data using the NEI model in {\tt XSPEC}. See Figure~3 (right) for the corresponding extraction slices for these fits. Abundances are fixed to Solar values. The column density is variable (top group) and fixed (bottom group) to the value obtained from region 2. Quoted uncertainties for 90\% C.L. for three (two) interesting parameter.
\end{table}

\begin{figure}
\includegraphics[width=0.5\textwidth]{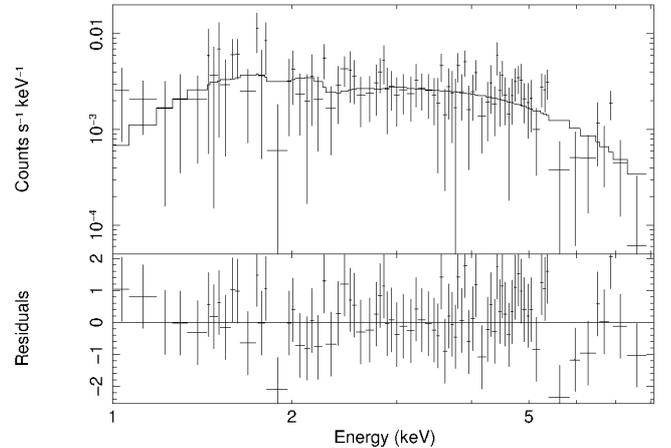}
 \caption{\textit{XMM-Newton} MOS 1 best-fit powerlaw (PL) model for region~1 (southern diffuse emission). The spectra have been rebinned for display purposes.}
\end{figure}

\begin{figure}
\includegraphics[width=0.5\textwidth]{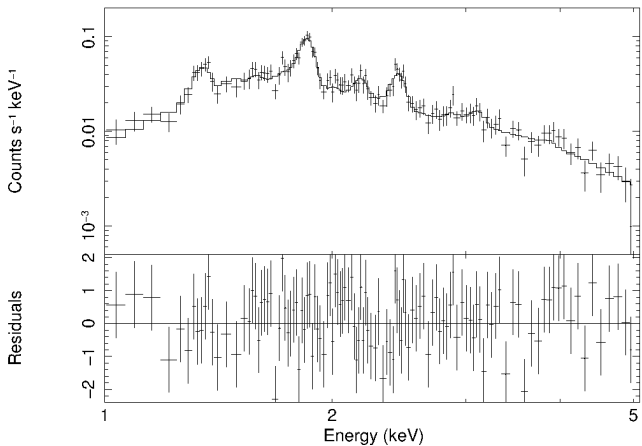}
 \caption{\textit{XMM-Newton} MOS 1 best-fit VPSHOCK model for region 2 (central diffuse emission). The spectra have been rebinned for display purposes.}
\end{figure}

\begin{figure}
\includegraphics[width=0.5\textwidth]{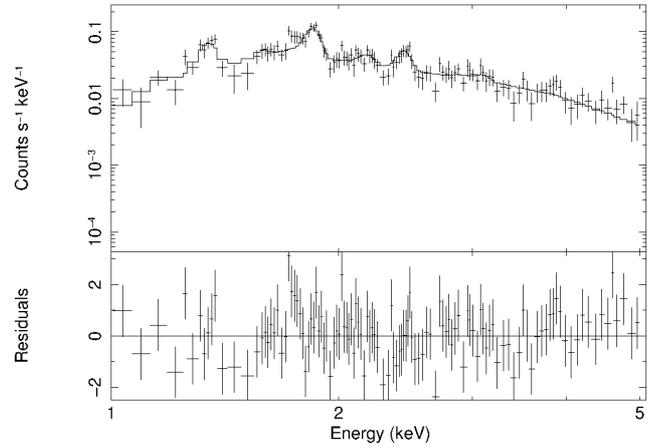}
 \caption{\textit{XMM-Newton} MOS 1 best-fit VSEDOV model for region 3 (global SNR). The spectra have been rebinned for display purposes.}
\end{figure}

\section{Discussion}
\label{5}

\subsection{Distance estimate}
\label{5.1}

The reported values for the distance to CTB 37B vary greatly. Using HI absorption measurements, Caswell et al. (1975) estimated the kinematic distance to be 10.2$\pm$3.5 kpc while Brand \& Blitz (1993) suggested a distance range of 5--9 kpc using a more recent galactic rotation curve model. Green (2009) estimates a distance of $\approx$8~kpc from HI absorption and Galactic rotation curve. Tian \& Leahy (2012) further suggest a distance of $\sim$13.2 kpc using the 1420 MHz radio continuum and 21 cm HI data from the Southern Galactic Plane Survey (SGPS). Surprisingly, most of those distance estimates are based on the same or comparable observations but with very different interpretation. Given the importance of the distance not only to our results but also to future studies, we present a short section on this subject to determine the appropriate and reliable distance estimates for these three remnants. 

All three SNRs (CTB 37A,  CTB 37B, and G348.5--0.0) of the CTB 37 complex show HI absorption up to the most negative velocity which still shows HI emission in their respective directions on the sky, as seen in the HI absorption profiles of Caswell et al. (1975) and Tian \& Leahy (2012). Therefore, they all must be located beyond the tangent point in this direction indicating a lower limit of 8.3~kpc for all three sources. This can be seen in Figure~6, where we show the radial velocity as a function of distance and Galacto-centric radius towards CTB 37B. For this, we assume a flat rotation curve for our Galaxy with the standard IAU endorsed values for the Galacto-centric radius of the sun ($R_\odot~=~8.5$~kpc) and the sun's orbital velocity ($\Omega_\odot = 220$ km\,s$^{-1}$).  This result is independent of the rotation curve used. This most negative velocity is about $-$120~km~s$^{-1}$ for CTB 37A and G348.5--0.0 (see HI absorption profiles published by Caswell et al. 1975 and Tian \& Leahy 2012). CTB~37B has an additional component at about $-$160~km~s$^{-1}$ in emission and absorption. We follow here the line of thought of Caswell et al. (1975), Frail et al. (1996), Reynoso \& Mangum (2000), and Aharonian et al. (2008) who also favour a location beyond the tangent point for CTB 37A and G348.5--0.0.

\begin{figure}
\includegraphics[width=0.5\textwidth]{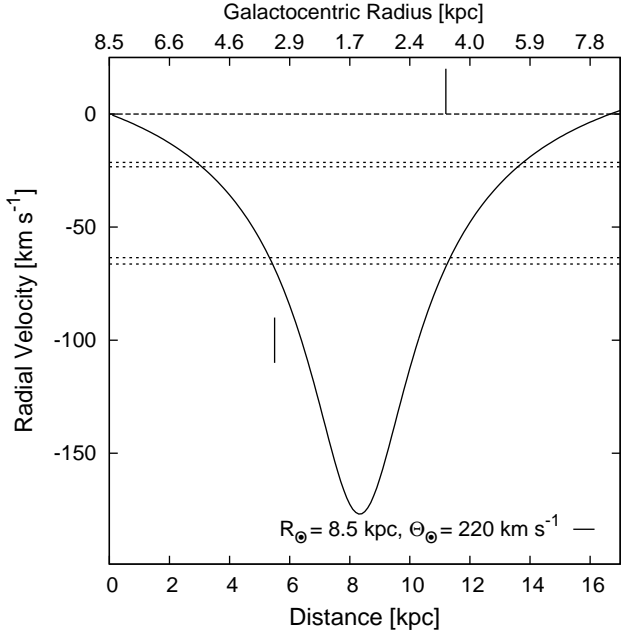}
 \caption{Radial velocity as a function of distance and Galacto-centric radius towards SNR CTB~37B (see Section 5.1 for details). For this, we assume a flat rotation curve for our Galaxy with the standard IAU endorsed values for the Galacto-centric radius of the sun ($R_\odot = 8.5$ kpc) and the sun's orbital velocity ($\Omega_\odot = 220$ km\,s$^{-1}$). The two vertical lines indicate the location of the near and far 3 kpc arm as determined by Dame and Thaddeus (2008). The dotted lines mark the velocity ranges of the masers and CO clouds related to CTB 37A and G348.5--0.0 (Frail et al. 1996).}
\end{figure}

Tian \& Leahy (2012) argue that there is CO emission towards CTB 37A and G348.5--0.0 at about $-$145~km~s$^{-1}$, and since these two sources do not show any absorption related to this molecular component they must be in front of it. But there is also no HI emission there, indicating that there might not be any or at least not enough atomic material that could be seen in emission or absorbing the continuum emission of CTB 37A and G348.5--0.0. This data was originally published by Reynoso \& Mangum (2000) and they did not mention any CO emission at this velocity range in their publication. In fact, their observation description indicates that the velocity of $-$145~km~s$^{-1}$ is at the edge of their observing band, which is why this looks cut off in Figure~2 of Tian \& Leahy (2012). This is probably just enhanced noise at the edge of the observed band rather than real CO emission.

A larger distance is supported by HI absorption shown for all three sources related to the velocity range of the near so-called 3~kpc arm found by Dame \& Thaddeus (2008). According to their CO data, the 3 kpc arm can be found towards a Galactic Longitude of 0 degrees at a radial velocity of $-$53.1 km~s$^{-1}$ and +56.0 km~s$^{-1}$ for the near and far part, respectively. Their fit to a longitude-velocity diagram gives a change of 4.16 km~s$^{-1}$degree$^{-1}$ for the near arm and 4.08 km~s$^{-1}$degree$^{-1}$ for the far arm. Extrapolated to our SNRs at about $-$11.5 degrees longitude, this results in central velocities of $-$101 km~s$^{-1}$ and +9 km~s$^{-1}$ for the near and far 3 kpc arms (see Figure 6). Both arms are supposed to be about 20 km~s$^{-1}$ wide (Dame \& Thaddeus 2008). Tian \& Leahy (2012) found a central velocity of $-$5 km~s$^{-1}$ for the far arm. Since the velocity of the near arm overlaps with the velocities of features in the inner Galaxy (Figure 6), it is difficult to distinguish between HI absorption features. But the CO emission towards CTB 37A and G348.5$-$0.0 in the velocity range around $-$100 km~s$^{-1}$ is likely related to the near 3 kpc arm (Reynoso \& Mangum 2000) and both sources show HI absorption for the entire velocity range of the CO emission. It was shown by Dame \& Thaddeus (2000) that there is HI emission related to the 3 kpc arm, therefore we would expect absorption for objects located behind it. These facts support a distance beyond the tangent point at 8.3 kpc for all the three SNRs.

CTB 37A and G348.5$-$0.0 are both related to maser emission at about $-$65 and $-$22 km/s, respectively (Frail et al. 1996). These systemic velocities for the two SNRs were confirmed by the detection of related CO(1-0) emission (Reynoso \& Mangum 2000) and CO(2-1) and CS(1-0) emission (Maxted et al. 2013) in the same velocity ranges. In the inner Galaxy, there is a distance ambiguity for negative radial velocities, but since we know that both SNRs are located beyond the tangent point, only the far distance is viable. A systemic velocity of $-$22 km/s would locate SNR G348.5$-$0.0 at a distance of 13.7 kpc using a flat rotation curve for the Milky Way Galaxy with the IAU endorsed values for the sun's Galacto-centric radius of $R_\odot$~=~8.5~kpc and its orbital velocity of $v_\odot$ = 220~km~s$^{-1}$ (see Figure 6). This would indicate a location beyond the far 3 kpc arm in this direction. Indeed, G348.5-0.0 is the only one out of the 3 SNRs showing absorption at positive velocities beyond the +6 km~s$^{-1}$, which could still be related to local features by velocity dispersion. There is a very clear absorption feature at about +15 km~s$^{-1}$ (Tian \& Leahy 2012) which falls in the +9 $\pm$ 10 km~s$^{-1}$ velocity range for the far 3 kpc arm at a distance of about 11.3~kpc (Dame \& Thaddeus 2008), which confirms a location beyond the 3 kpc arm at 13.7 kpc. CTB 37A does not show any absorption in this velocity range and therefore must be located in front of the far 3 kpc arm. Its radial velocity would translate to a distance of 11.3 kpc, similar to the far 3 kpc arm, using the flat rotation curve and it therefore must be located somewhat closer than the far 3 kpc arm at about 11 kpc.

CTB 37B also does not show an absorption feature in the velocity range beyond +6 km~s$^{-1}$ and therefore must be located in front of the far 3 kpc arm. Since we do not have any additional information about its systemic velocity and there has not been any discovery of related molecular gas that would help to put further limitation on its distance, we can only give a lower limit of 8.3 kpc for the tangent point and an upper limit of 11.3 kpc for the far 3 kpc arm. Therefore, the distance to CTB 37B results in 9.8$\pm$1.5 kpc and in this study, we adopt a value of $D_{9.8}$ $\sim$ 9.8 kpc.

\subsection{Radio properties of SNR CTB 37B}
\label{5.2}

There have been numerous radio continuum observations of CTB~37B in the past with both, single antenna radio telescopes and interferometers. However, the determination of reliable flux densities and resulting spectral index have been challenging due to the complexity of this area on the Galactic plane combined with either low resolution observations or observations with an interferometer that lack short spacing information. This can easily be seen in the spectrum published in Figure~6 of Kassim et al. (1991). They found a spectral index of $\alpha = -0.3$, but there is an extremely large scatter of flux densities at every frequency. In particular, there are 3 flux densities displayed for 5 GHz the highest of which is more than twice the lowest. It is also unclear whether to include a bridge of radio emission (Kassim et al. 1991), which is extending from under the shell-type part of the SNR to the south-east. Under those circumstances, reliable flux density determinations are difficult and therefore, spectral index determinations are highly uncertain. We require high resolution radio continuum observations with a high spatial dynamic range to determine reliable flux densities. Future radio continuum surveys with Square Kilometre Array pathfinder telescopes (Norris et al. 2013) will provide the observations necessary to disentangle this whole complex area and we can determine reliable spectra for all sources.

\begin{figure}
\includegraphics[width=0.5\textwidth]{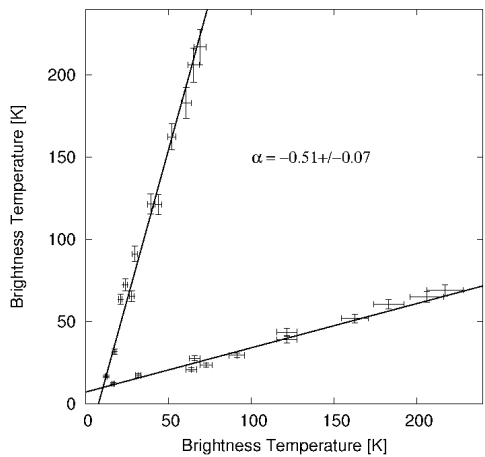}
 \caption{TT-plot to determine the spectral index of CTB 37B between 843~MHz and 1420~MHz (see section 5.2 for details). The two straight lines are linear functions from least-square fits to the radial profile at 843 MHz as a function of 1420 MHz (left) and vice versa (bottom), resulting in an averaged spectral index of $\alpha = $-$0.51 \pm 0.07$. The derived spectral indices from the individual fits are $\alpha$ = $-$0.49 $\pm$ 0.10 and $\alpha$ = $-$0.52 $\pm$ 0.10.}
\end{figure}

We attempted to establish a spectral index for the bright shell with a different method using the only reliable data easily available, which is the 1420 MHz Southern Galactic Plane Survey observation (SGPS; Haverkorn et al. 2006) and the data from the 843 MHz SNR catalogue observed with the MOST (Whiteoak \& Green 1996). To determine the radio spectral index of the bright shell-type feature covered by our X-ray observations, we used the so-called ``TT-Plot'' method described by Turtle et al. (1962). The amplitudes at both frequencies are  radial profiles calculated from the geometric centre of the almost circular shell at  RA (J2000) = 17$^h$ 13$^m$ 52.5$^s$ and DEC (J2000) = $-$38$^\circ$ 11$'$ 30$"$. Those are based on ring-averaged brightness temperatures with a ring-width of 0.5$'$. The higher resolution 843 MHz map was convolved to the resolution of the SGPS data before the radial profile was determined. The result is shown in Figure~7. TT-plots give reliable results for spectral indices of bright objects on top of diffuse extended emission. This diffuse emission would produce an offset from the origin of the diagram for the TT-plot fit if its spectral index is different. In this case, the offset is positive for the higher frequency axis, indicating a flatter spectrum for the diffuse component. This diffuse emission bridge is likely unrelated thermal emission. From the TT-plot, we fit a spectral index of $\alpha$ = $-$0.51$\pm$0.07 for the bright shell, which indicates a Sedov phase shell-type SNR.

\subsection{X-ray properties of CTB 37B}
\label{5.3}

We derive the X-ray properties of CTB 37B using the VSEDOV model parameters summarized in Table 3. The radius of the SNR is taken as $\sim$5$\arcmin$.1, as estimated by Aharonian et al. (2008), which translates to a physical size of $R_s$ = 14.5$D_{9.8}$ pc = 4.5$\times$10$^{19}$ $D_{9.8}$ cm. The volume of the SNR shell, estimated by assuming that the plasma fills a sphere, is given by $V$ = 3.8$\times$10$^{59}$~cm$^3$. The emission measure $EM$~=~$\int n_en_HdV$ $\sim$ $fn_en_HV$ is defined as a measure of the amount of plasma available to produce the observed flux, where $n_e$ is the post-shock electron density for a fully ionized plasma, $n_H$ is the mean Hydrogen density, and $f$ is the volume filling factor of the hot gas. The emission measure can be estimated from the normalization $K$~=~($10^{-14}$/4$\pi$$D^2$)$\int$$n_en_HdV$ (obtained from the X-ray spectral fits) as EM~=~7.8$^{+0.7}_{-0.6}$$\times$10$^{57}$$D_{9.8}^{2}$~cm$^{-3}$. For cosmic abundance plasma and the strong shock Rankine-Hugoniot jump conditions, the ambient density $n_0$ can be estimated from the electron density $n_e$~=~4.8$n_0$. Using the above equations, we calculated the post-shock electron density as $n_e$~=~0.16$\pm$0.01~$f^{-1/2}$$D_{9.8}^{-1/2}$ cm$^{-3}$ and the ambient density as $n_0$~=~0.03$\pm$0.01 $f^{-1/2}$$D_{9.8}^{-1/2}$ cm$^{-3}$ of the X-ray emitting plasma. The low inferred ambient density suggests that the SNR is mostly expanding into a low-density medium, possibly a bubble blown by the supernova progenitor. 

A lower or upper estimate on the remnant's age can be inferred by assuming a free expansion or Sedov phase of evolution, respectively. For an initial expansion velocity of $v_0$~$\sim$~5000~km~s$^{-1}$ (Reynolds 2008), the SNR's size implies a free expansion age of $\gtrsim$2800~$D_{9.8}$~yr. The swept-up mass is calculated as $M_{sw}$~=~($\frac{4}{3}\pi$$R_s^3$)$\times$1.4$m_p$$n_0$ = (15$\pm$5)~$f^{-1/2} D_{9.8}^{5/2}$~M$_{\sun}$, under the assumption of a uniform ambient density medium. The Sedov age of the SNR is given by $t_{SNR}$ = $\eta$$R_s$/$V_s$ where $\eta$ = 0.4 for a blast wave expansion in the Sedov phase (Sedov 1959). The shock velocity can be estimated as $V_s$ = (16$k_BT_s$/3$\mu$$m_H$)$^{1/2}$ assuming full equilibration between the electrons and ions (Sedov 1959), where $\mu$ = 0.604 is the mean mass per free particle for a fully ionized plasma, $k_B$~=~1.38$\times$10$^{-16}$ ergs~K$^{-1}$ is Boltzmann's constant. The post-shock temperature $T_s$ is related to the temperature inferred from the X-ray fit ($T_X$) as $T_X$ $\sim$ 1.27$T_s$ in the Sedov phase (since the temperature increases inward behind the shock radius; Rappaport et al. 1974).  We infer a shock velocity $V_s$~=~915$\pm$70~km~s$^{-1}$ for the forward shock and a Sedov age $t_{SNR}$~=~(6.2$\pm$0.5)$\times$10$^3$$D_{9.8}$ yr for CTB~37B. The shock age (time since the passage of shock), inferred from the ionization timescales ($n_et$) is given by $t$ = (1.2$\pm$0.1)$\times$10$^4$~$f^{1/2}$$D^{1/2}_{9.8}$~yr.  The SNR explosion energy can be estimated based on the Sedov blast wave model in which a supernova with an explosion energy $E$, expands into an ISM of uniform density $n_0$ as $E$~=~$\frac{1}{2.02}$$R_s^5$$m_n$$n_0$$t_{SNR}^{-2}$ = (1.8$\pm$0.6)$\times$10$^{50}f^{-1/2}D^{5/2}_{9.8}$ ergs, where $m_n$~=~1.4$m_p$ is the mean mass of the nuclei and $m_p$ is the mass of the proton. If full electron-ion equilibration has not been achieved in the shock for the Sedov phase, the shock velocity $V_s$ determined from the electron temperature will be a lower limit to the shock temperature $T_s$ leading to an underestimation of the remnant's explosion energy. 

As summarized in Table 3, the southern side of the remnant shows the presence of nonthermal X-ray emission with a hard photon index of $\Gamma$ = 1.3$\pm$0.3, consistent with the results obtained for this region by Nakamura et al. (2009). Frail et al. (1996) reported the evidence of significant OH maser emission near the CTB 37 complex, indicating dense shock-heated gas. Although no maser emission has been confirmed from the direction of CTB 37B, this hard nonthermal emission which appears too hard for a shock acceleration interpretation could possibly be due to the presence of a molecular cloud to the east that is interacting with the nearby remnant CTB~37A (Reynoso \& Mangum 2000). The bright X-ray emission seen on the eastern side of the remnant, together with the fact that this is in a complex region of the sky and the source was detected with Fermi (Xin et al. 2016), it is reasonable to assume that the X-rays could arise from molecular interaction. However, the X-ray photon index is harder than expected from such a scenario. Recently, Tanaka et al. (2018) discovered nonthermal bremsstrahlung emission from SNR~W49B, which also shows a very hard photon index of $\Gamma$ = 1.4$^{+1.0}_{-1.1}$ in the 10--20 keV range. Alternatively, the diffuse emission from the southern part of CTB 37B could also be an unrelated PWN, since the hard photon index obtained from this region is more consistent with photon indices observed for PWNe although on the hard end (e.g., Gotthelf 2003; Kargaltsev \& Pavlov 2008). In order to check for any candidate neutron star, we revisited the point sources that were removed from region 1 to study the SNR emission and cross-referenced with the \textit{XMM-Newton} serendipitous source catalogue (3XMM-DR8\footnote{http://xmm-catalog.irap.omp.eu}). We find two point sources, 3XMM~J171410.5--381446 and 3XMM~J171409.6--381505, with coordinates matching the removed sources from region~2.  However, deeper Chandra observations are required to confirm the nature of these sources and the hard emission to the south of the magnetar.

\subsection{Implications for the magnetar CXOU J171405.7-381031}

The spin-evolution of CXOU J171405.7--381031 is erratic and varies in time. This makes the characteristic pulsar (PSR) age, $\tau_\text{PSR}$, an unreliable estimate of its true age. Instead, following Gao et al. (2014, 2015) and Rogers \& Safi-Harb (2017), we use the estimated SNR age to approximate the neutron star (NS) age. This allows us to put limits on the calculated properties of the NS such as magnetic field strength, for example.

The standard definition of the NS braking index is given in terms of the spin-period and its derivatives, 
\begin{equation}
n=2-\frac{P\ddot{P}}{\dot{P}^2}.    
\end{equation}
The standard magneto-dipole braking scenario assumes a constant braking index, $n=3$, as well as constant spin-period and derivatives. However, any changes in the spin mean that $n$ must also change over time. From a practical point of view, it is extremely difficult to use this expression directly since it involves higher derivatives of the spin period. These period derivatives are small and sensitive to rotational irregularities. Instead of this expression, it is more useful to express the characteristic age of the NS in terms of the braking index,
\begin{equation}
\tau_\text{PSR} = \frac{P}{(n-1)\dot{P}} \left[ 1-\left( \frac{P_0}{P} \right)^{(n-1)}\right]
\label{tauPSR}
\end{equation}
where $P$ and $\dot{P}$ are the observed period and the first period derivative. Here we assume that the characteristic age should be well-approximated by the age of CTB 37B. Thus, any constraint on the age of the SNR simultaneously provides a constraint on the evolution of the associated magnetar CXOU J171405.7--381031.

If the torque acting to brake the rotation of the star can be treated as constant over the life of the magnetar, we can then write 
\begin{equation}
\frac{P_0}{P} = \left[ 1 - (n-1)\tau_\text{PSR} \frac{\dot{P}}{P} \right]^\frac{1}{(n-1)}.
\end{equation}
The braking index is $n=3$ for magneto-dipole braking, and $n=5$ for braking by gravitational wave emission. Thus, we set the characteristic age of the magnetar equal to the upper and lower age estimates of CTB 37B, $\tau_\text{PSR} = \tau_{\text{SNR}\pm}$, to find limits for the possible value of the braking index in terms of the ratio of initial to observed period, $P_0/P$. This plot is shown in Figure 8. However, the assumption of a constant braking index is unrealistic, particularly given the variable nature of the rotational evolution of the magnetar. We are assuming that the measured variables $P$ and $\dot{P}$ are also constant and provide a good estimate of the persistent spin-parameters. Thus, the best this calculation can provide is an approximation to the true average braking index and initial spin-period.

\begin{figure}
\includegraphics[width=0.5\textwidth]{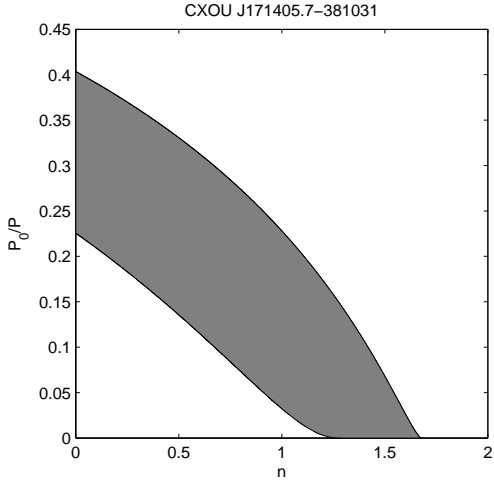}
 \caption{Relationship between the ratio of initial and observed spin periods $P_0/P$ and braking index $n$. The grey region represents combinations for which $\tau_{\text{SNR}-} \leq \tau_\text{PSR} \leq \tau_{\text{SNR}+}$, where $\tau_{\text{SNR}\pm}$ are the upper and lower limits of the SNR age. This plot shows that for initial spin periods below $0.4$ the observed value, the braking index $n<1.68$ is required if it remains constant over the life of the magnetar.}
\end{figure}

A more realistic scenario for the dynamical evolution of magnetars is a time-varying braking index that depends on the decay of the magnetic field. The condition that the age of the magnetar and the age of CTB 37B agree imposes a strong constraint on models of magnetic field decay. For a time-dependent magnetic field we have 
\begin{equation}
P\dot{P} = b B(t)^2 
\end{equation}
where $b$ is a constant of proportionality. This results in a time-varying braking index, 
\begin{equation}
n=3-2\frac{\dot{B}}{B}\frac{P}{\dot{P}}
\end{equation}
For any type of decaying field, regardless of the details of the model parameterization, $\dot{B}<0$ and $n>3$. A consequence of this parameterization in terms of the elapsed time, $t$, means that many of the properties of the star are time dependent. For example, $t$ represents the true age of the NS and must therefore match the SNR age. This means that the characteristic age is now considered a dynamical quantity that varies with time and can differ from the SNR age over the course of the NS' life. Generally $\tau_\text{PSR} \neq t$, such that field decay causes neutron stars to appear older than they actually are. This effect was used by Nakano et al. (2015) to successfully describe the discrepancy between the age of the anomalous X-ray pulsar (AXP) 1E 2259+586 and its host SNR, CTB 109. The field decay model of Nakano et al. (2015) use the magnetic field parameterization 
\begin{equation}
\frac{\text{d}B}{\text{dt}}=-aB^{1+\alpha}
\end{equation}
where $a$ is a positive constant, and $\alpha$ the field decay index. In Figure 9, we show the field decay models found by Nakano et al. (2015). The three curves show field decay index $\alpha=1.4$, $1.0$, and $0.6$, and all produce slightly different evolutionary behavior. An attractive feature of the field decay approach is that in addition to 1E 2259+586, the models simultaneously describe the relationship between the ages of AXP 1E 1841--045 and its associated SNR, Kes~73. However, the age bounds on CTB 37B reported in this work compared to the characteristic age of CXOU J171405.7--381031 argue against a decaying magnetic field to describe this object. Instead, the relationship between SNR and NS characteristic ages are reversed, and the SNR looks older (2.8--6.2 kyr) than the apparently young (1.0 kyr) magnetar that it hosts.

One particularly well-known mechanism for generating a magnetar braking index $1<n<3$ is through the emission of a relativistic wind (Gao et al. 2015). Using the period and period derivative of CXOU J171405.7--381031, we use the equatorial dipole field $B_d$ = 3.2$\times$10$^{19} \sqrt{P \dot{P}}$ as an upper limit. If the magnetar spins down with a low braking index the corresponding luminosity of the emitted particle wind is 
\begin{equation}
    L_\text{w} = (3-n)^2\frac{6 I^2 c^3}{B_\text{d}^2 R^6} \left(\frac{\dot{P}}{P}\right)^2
\end{equation}
where $c$ is the speed of light in cm/s, the moment of inertia is taken to be $I=10^{45}$ g cm$^2$, and the radius of the neutron star is $R=10$~km. Using upper and lower braking index estimates of 1 $\leq n \leq$ 2.5, we find a range of wind luminosity between 4.5$\times$10$^{34}$ ergs~s$^{-1}$ and 7.2$\times$10$^{35}$ ergs~s$^{-1}$. To compare with the energetics of CXOU~J171405.7--381031, we calculate the luminosity of the magnetar given its flux in the range (1.2--1.6) $\times$ 10$^{-12}$ ergs~cm$^{-2}$~s$^{-1}$ (Halpern \& Gotthelf 2010b). Using the average, 1.4$\times$10$^{-12}$~ergs~cm$^{-2}$~s$^{-1}$, and the upper and lower limit of the SNR distance $D$ = 9.8$\pm$1.5~kpc, we find a range of luminosities between (1.6--2.1) $\times$ 10$^{34}$ $D_{9.8}$ ergs~s$^{-1}$. Thus, we conclude that the emission of a relativistic wind may be one way to account for the energetics of the neutron star, provided that the age constraint is simultaneously obeyed. Other alternative scenarios which produce a time-varying braking index $n<3$ and allow the characteristic age of a neutron star to appear smaller when compared to its true age are discussed in Rogers \& Safi-Harb (2016, 2017) and references within.

\begin{figure}
\includegraphics[width=0.5\textwidth]{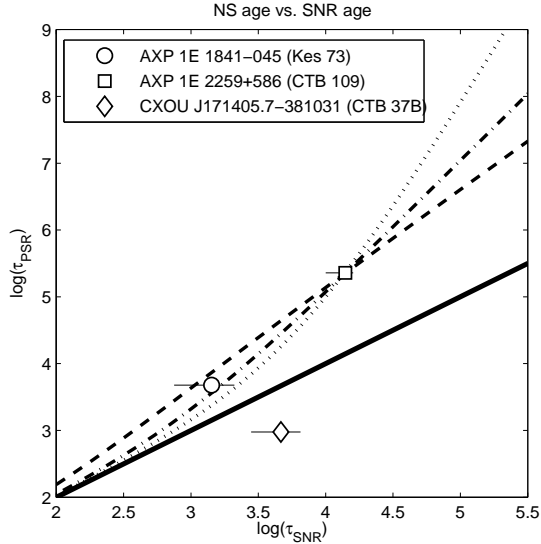}
\caption{Characteristic PSR age $\tau_\text{PSR}$ as a function of SNR age $\tau_\text{SNR}$. The dashed, dash-dotted and dotted lines represent the three decaying field solutions found in Nakano et al. (2015) to match the SNR and PSR ages of 1E 2259+586 and CTB 109. The solutions have field decay index $\alpha = 1.4$, $1.0$ and $0.6$, respectively. While the field decay model works well for both AXPs 1E 2259+586 and 1E 1841--045, the refined age for CTB 37B presented in this work argues against the explanation of field decay to resolve the discrepancy with the age of CXOU J171405.7--381031.}
\label{fig:Bdecay}
\end{figure}

The conclusive test of the field decay scenario would be a measurement of the actual braking index of CXOU J171405.7--381031, which would rule out field decay as an explanation for the energetics of the neutron star provided $n<3$, and argue for the observed age relationship $\tau_\text{SNR}>\tau_\text{PSR}$ found in this work. Unfortunately, a precise measurement of this quantity is not possible given the erratic nature of the magnetar. Thus, we are left with only estimates of the age and braking index which rely on assumptions about the average values of the parameters in the problem.

\section{Conclusion}

We have performed a spatially resolved spectroscopic study of the diffuse emission regions within SNR CTB 37B using \textit{Chandra} and \textit{XMM-Newton}. In addition, we used the radio data to constrain the distance to the remnant and obtain $D$~=~9.8$\pm$1.5~kpc. At both the radio and X-ray wavelengths, the eastern side of the SNR appears brighter than the western side. The spectral analysis confirms that the diffuse emission from within the remnant is best described by a one-component thermal model showing solar or sub-solar abundances. The ionization timescales suggest that the plasma has not reached ionization equilibrium. The overall imaging and spectral properties of CTB~37B favor the interpretation of a remnant $\sim$2.8--6.2 kyr old, propagating in a low-density medium under the assumption of a Sedov evolutionary phase. For a distance of 9.8~kpc, we infer a shock velocity of $V_s$ = 915$\pm$70~km~s$^{-1}$ and explosion energy of $E$~=~(1.8$\pm$0.6)$\times$10$^{50}$ ergs. The diffuse emission towards the southern part of CTB~37B is represented by a hard nonthermal emission whose nature is puzzling, but could be due to shock with a nearby molecular cloud (that is yet to be detected) or likely originating from a PWN. Deep observations are needed to explain the nature of this intriguing hard X-ray emission.

\section*{Acknowledgments}
This research made use of NASA's Astrophysics Data System (ADS) and of HEASARC maintained at NASA's Goddard Space Flight Center (GSFC). We thank the referee for comments that helped improve the clarity of the paper. HB is supported by the NSF award number 1516512. S. Safi-Harb acknowledges support from the Natural Sciences and Engineering Research Council of Canada (NSERC) through the Canada Research Chairs and the Discovery Grants Programs. E.V.G. acknowledges support from NASA grant 80NSSC18K0452.

\end{document}